\documentclass[aps,prass,superscriptaddress,twocolumn]{revtex4}
\usepackage{graphicx}
\usepackage{amsmath}
\usepackage{color}
\usepackage{bm}

\newcommand{\ket}[1]{\ensuremath{\left|{#1}\right\rangle}}
\newcommand{\bra}[1]{\ensuremath{\left\langle{#1}\right |}}

\newcommand{\beq}{\begin{equation}}
\newcommand{\eeq}{  \end{equation}}
\newcommand{\bea}{\begin{eqnarray}}
\newcommand{\eea}{  \end{eqnarray}}
\newcommand{\bit}{\begin{itemize}}
\newcommand{\eit}{  \end{itemize}}

\graphicspath{%
    {converted_graphics/}% inserted by PCTeX
    {C:/Users/PHSR/Desktop/Followup/FINALX/NewFigs/}% inserted by PCTeX
}

\begin{document}

\title{Flow of Quantum Correlations from a Two-Qubit System to its Environment}
\author{G. H. Aguilar}
\affiliation{Instituto de F\'{\i}sica, Universidade Federal do Rio de Janeiro, Caixa
Postal 68528, Rio de Janeiro, RJ 21941-972, Brazil}
\author{O. Jim\'enez Far\'{\i}as}
\affiliation{Instituto de F\'{\i}sica, Universidade Federal do Rio de Janeiro, Caixa
Postal 68528, Rio de Janeiro, RJ 21941-972, Brazil}
\author{A. Vald\'es-Hern\'andez}
\affiliation{Instituto de F\'{\i}sica, Universidade Federal do Rio de Janeiro, Caixa
Postal 68528, Rio de Janeiro, RJ 21941-972, Brazil}
\affiliation{Instituto de Fisica, Universidad Nacional Autonoma de Mexico, A.P. 20-364, Mexico D.F., Mexico}
\author{P. H. Souto Ribeiro}
\affiliation{Instituto de F\'{\i}sica, Universidade Federal do Rio de Janeiro, Caixa
Postal 68528, Rio de Janeiro, RJ 21941-972, Brazil}
\author{L. Davidovich}
\affiliation{Instituto de F\'{\i}sica, Universidade Federal do Rio de Janeiro, Caixa
Postal 68528, Rio de Janeiro, RJ 21941-972, Brazil}
\author{S. P. Walborn}
\affiliation{Instituto de F\'{\i}sica, Universidade Federal do Rio de Janeiro, Caixa
Postal 68528, Rio de Janeiro, RJ 21941-972, Brazil}

\date{\today }

\begin{abstract}
The open-system dynamics of entanglement plays an important role in the assessment of the robustness of quantum information processes and also in the investigation of the classical limit of quantum mechanics. Here we show that, subjacent to this dynamics, there is a subtle flow of quantum correlations. We use a recently proposed optical setup, which allows joint tomography of system and environment, to show that the decay of an initial bipartite entangled state leads to the build up of multipartite  entanglement and quantum discord, the latter  exhibiting a non-analytic behavior that signals the emergence of maximal genuine quantum entanglement. The origin of this analyticity is shown to be distinct from similar behavior previously found in bipartite systems. Monogamy relations within the context of open-system dynamics explain this new phenomenon. 
\end{abstract}

\maketitle
%\preprint{}

%Title of paper
%\title{Experimental investigation of the transition from bipartite to genuine multipartite entanglement in %three qubit systems}

%--------------------------------------

% insert suggested PACS numbers in braces on next line

% insert suggested keywords - APS authors don't need to do this
%\keywords{}
%%%%%%%%%%%%%%%%%%%%%%%%%%%%%%%%%%%%%%%%%%%%%%%%%%%%%%%%%%
\section{Introduction}
%%%%%%%%%%%%%%%%%%%%%%%%%%%%%%%%%%%%%%%%%%%%%%%%%%%%%%%%%%

The existence of quantum correlations is a fundamental difference between quantum and classical physics. 
Entanglement is one kind of quantum correlation that is stronger than all existing classical correlations. 
Its understanding has led to the development of communication protocols like quantum teleportation \cite{bennett93} and quantum cryptography \cite{ekert91}, and enables measurements with a precision that exceeds the standard quantum limit \cite{giovannetti04}. For Quantum Information Science, entanglement is a valuable resource\cite{chuang00}. However, it has been  demonstrated that not all the quantum correlations are captured by entanglement \cite{henderson01, ollivier02}. A different class of correlations called ``Quantum Discord" (QD) has also been considered as a resource for non-classical computation,
namely in the DQC1 model \cite{knill98, datta05}. 
An experimental implementation of this model was performed in an optical setup \cite{lanyon08} and also in the context of Nuclear Magnetic Resonance \cite{passante11}. Moreover, it was recently demonstrated that the fidelity for remote preparation of states, a variation of the teleportation protocol, is related to a measure of QD \cite{dakic12}. However, the scope and computational power of quantum discord is still a matter of debate. 
 
In general, physical systems containing quantum correlations cannot be completely isolated from the environment. In fact, these systems always interact with their surroundings, leading to noise and decoherence \cite{giulini87,zurek03}. The investigation of this interaction between system and environment has led to the observation of interesting phenomena, like the sudden death of entanglement \cite{yu04,almeida07, salles08}, and sudden changes of quantum discord  \cite{maziero09, xu10}. It has also been useful in the development of strategies to protect the quantum correlations against environmental interactions,
and to investigate fundamental problems like the border between the quantum and the classical world \cite{giulini87,zurek03,blume-kohout06,zurek09,cornelio12}. 

Since the degrees of freedom of the environment are unaccessible, they are usually ignored. However, if one considers the complete system including the environment, additional useful information can be obtained about the dynamics of entanglement \cite{salles08,farias12b} in the overall system, which may be useful to the investigation of the quantum-classical transition \cite{zurek09,farias12b}. 

%Indeed, the imprint of the system into its environment helps to characterize the so-called pointer 
%states \cite{blume-kohout06,zurek09,cornelio12}. Among the states of the system, they are more immune to decoherence, and have been shown to imprint redundant information onto the environment \cite{blume-kohout06,zurek09}. In this way, one does not need to probe the whole surroundings of the system for its characterization. This result, which unveils a subtle property of the classical world, motivates our study of the joint entanglement dynamics of a system and its environment.
%In particular, when each part of an entangled system interacts with its own environment, and there is no interaction between the parts, the dynamics of system plus environment involves only local unitary evolutions. Therefore, one should expect a redistribution of the initial entanglement between system and environment. 
\par
\begin{figure}[tbp]
\centering 

  \includegraphics[width=8cm]{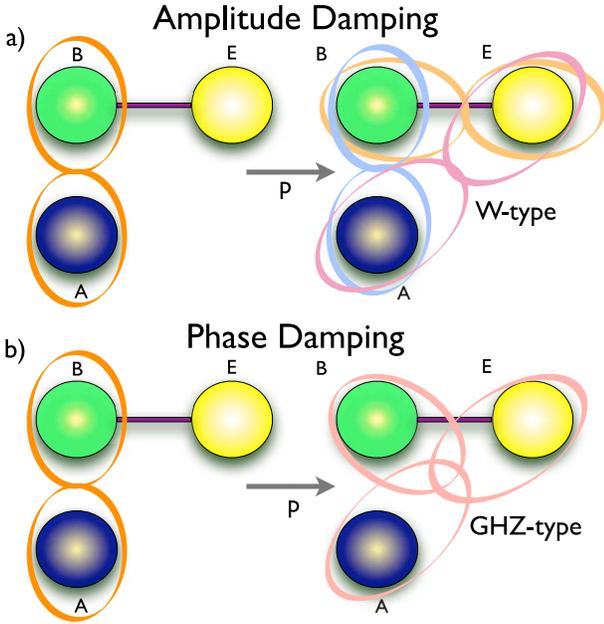}

\caption{Evolution of entanglement. 
In a) system $B$ is submitted to an AD channel. The initial entanglement is redistributed 
in all bipartitions, so that only bipartite or W-type of tripartite entanglement is generated. 
In b) system $B$ is submitted to a PD channel. In this case tripartite entanglement is generated.}
\label{fig:evolution}
\end{figure}

%In this paper, we study theoretically and experimentally
%the evolutions sketched in Fig. \ref{fig:evolution}. Systems $A$
%and $B$ are qubits initially entangled, whereas $E$, the local environment of $B$,
%is in its ground state. At $t=0$, $B$ starts to interact with $E$ in such a
%way that along the evolution, the initial qubit-qubit
%entanglement flows to genuine tripartite entanglement, 
%which may be of GHZ type, $\alpha|000\rangle+\beta|111\rangle$, 
%or of the W type, $\alpha|100\rangle+\beta|010\rangle+\gamma|001\rangle$, 
%depending on the specific interaction between $B$ and $E$  as well as the initial state of the bipartite system. With a novel and all-optical
%experimental setup that was recently presented in \cite{farias12b}, and allows
%the complete control and measurement of the environmental degrees of freedom, 
%we investigate the distribution of entanglement
%for two of the most common quantum channels: amplitude damping (AD) and
%phase damping (PD)\cite{chuang00}. In this setup, the system $AB$ corresponds to two polarization-entangled photons, while the environment is taken as the spatial modes of one of the photons.  The interactions considered here are such that the environment can have at most one excitation, so it can  be considered as a qubit. The two states of $E$ correspond to the two spatial modes of one of the photons (the one encoding system $B$).
%%%%%%%%%%%%%%%%%%%%%%%%%%%%%%%%%%%%%%%%%
%%%%%%%%%%%%%%%%%%%%%%%%%%%%%%%%%%%%%%%%%%
In this paper, we use a novel and all-optical experimental set-up that was recently presented in \cite{farias12b} and allows the complete tomography of a two-qubit system and its environment. Fig. \ref{fig:evolution} illustrates the relevant evolutions. Systems $A$
and $B$ are qubits initially entangled, whereas $E$, the local environment of $B$,
is in its ground state. While \cite{farias12b} concentrated on the experimental observation of the emergence of multipartite entanglement between system and environment, here we demonstrate subtle aspects of this process, which are analyzed theoretically and experimentally. In particular, we: i)derive  a simple expression that quantifies the emergence of a GHZ type of entanglement, given by a product of the initial tangle and a function that depends solely on the Kraus operators describing the non-unitary dynamics;
ii) demonstrate that the genuine multipartite quantum discord signals the development of genuine multipartite entanglement; iii) unveil and explain a novel non-analytical behavior of the genuine multipartite discord,  which is shown to signal the appearance of maximal W-type entanglement; iv) introduce a new method to analyze experimental results concerning quasi-pure states, which allows one to compare data obtained from weakly-noisy systems with the theoretical results conceived for pure states; and v) observe and explain the appearance of genuine GHZ entanglement due to experimental noise. 

%%%%%%%%%%%%%%%%%%%%%%%%%%%%%%%%%%%%%%%%%
%%%%%%%%%%%%%%%%%%%%%%%%%%%%%%%%%%%%%%%%

%This article is organized as follows: in Section \ref{sec:theory},
%we develop a theoretical framework for describing the dynamics of tripartite pure states 
%(involving $A$, $B$ and $E$). We show that the decay of the initial two-qubit entanglement may lead to the emergence of genuine tripartite entanglement in the total $ABE$ system. A detailed explanation of the experimental setup is introduced in section \ref{sec:experiment}. 
%The experimental results are presented in section \ref{results1}.
%The conclusions and perspectives are summarized in Sec. \ref{sec:conclusions}.

%This article is organized as follows: in Section \ref{sec:theory},
%we develop a theoretical framework for describing the open-system dynamics of quantum correlations.  A detailed explanation of the experimental setup is introduced in Section \ref{sec:experiment}. 
%The experimental results concerrning the flow of entanglement and quantum discord are presented in Section \ref{results1}.
%The conclusions and perspectives are summarized in Sec. \ref{sec:conclusions}.

%%%%%%%%%%%%%%%%%%%%%%%%%%%%%%%%%%%%%%%%
%%%%%%%%%%%%%%%%%%%%%%%%%%%%%%%%%%%%%%%
This article is organized as follows: in Section \ref{sec:theory},
we develop a theoretical framework for describing the open-system dynamics of entanglement.  A detailed explanation of the experimental setup  is introduced in Section \ref{sec:experiment}. Here we propose an new way of treating the experimental data in order to analyze quasi pure states.
%, which agrees correctly with the theory for pure states. 
In Section \ref{results1}, the experimental results concerning the flow of entanglement are discussed in terms of the quantum discord for the explanation of new non analytical behavior of the quantum correlations. The conclusions and perspectives are summarized in Section \ref{sec:conclusions}.

%%%%%%%%%%%%%%%%%%%%%%%%%%%%%%%%%%%%%%%%%%%%%%%%%%
%\section{Dynamics of the distribution of entanglement}
%%%%%%%%%%%%%%%%%%%%%%%%%%%%%%%%%%%%%%%%%%%%%%%%%%%%%%
\section{Theory}
\label{sec:theory}
%%%%%%%%%%%%%%%%%%%%%%%%%%%%%%%%%%%%%%%%%%%%%%%%%%%%%%%%%%
We assume that the initial state of the complete ($ABE$) system is given by 
\begin{equation}
\left\vert {\varphi (0)}\right\rangle \!=\left[ \alpha \left\vert {11}%
\right\rangle +\beta \left\vert {10}\right\rangle +\gamma \left\vert {01}%
\right\rangle +\delta \left\vert {00}\right\rangle \right] _{AB}\left\vert {0%
}\right\rangle _{E},  \label{psi0}
\end{equation}%
where $|\alpha |^{2}+|\beta |^{2}+|\gamma |^{2}+|\delta |^{2}=1$. At time $%
t=0$, system $B$ begins to interact with its environment $E$ according to a
unitary transformation $U_{BE}(t)$ including both systems. Since the initial
state of $E$ is $\left\vert 0 \right\rangle$, and 
only two states of the environment are involved in the dynamics, 
only two Kraus operators are needed to describe the evolution, namely $%
\hat{K}_{0}=\left\langle {0}\right\vert _{E}U_{BE}\left\vert {0}%
\right\rangle _{E},$ and $\hat{K}_{1}=\left\langle {1}\right\vert
_{E}U_{BE}\left\vert {0}\right\rangle _{E}$\cite{salles08}. Thus, the unitary
evolution of the system $BE$ can be represented by the map: 
\begin{eqnarray}
\left\vert {0}\right\rangle _{B}\left\vert 0\right\rangle _{E} &\rightarrow &%
\hat{K}_{0}(t)\left\vert {0}\right\rangle _{B}\left\vert {0}\right\rangle
_{E}+\hat{K}_{1}(t)\left\vert {0}\right\rangle _{B}\left\vert {1}%
\right\rangle _{E},  \label{Map} \\
\left\vert {1}\right\rangle _{B}\left\vert 0\right\rangle _{E} &\rightarrow &%
\hat{K}_{0}(t)\left\vert {1}\right\rangle _{B}\left\vert 0\right\rangle _{E}+%
\hat{K}_{1}(t)\left\vert {1}\right\rangle _{B}\left\vert 1\right\rangle _{E},
\notag
\end{eqnarray}%
where each $\hat{K}_{i}$ acts only on the vectors of system $B$. Writing $\hat{K}%
_{0}$ and $\hat{K}_{1}$ as (we omit the time dependence of the matrix elements) 

\label{krausmap}
\begin{equation}
\hat{K}_{0}=\left( 
\begin{array}{ll}
m_{00} & m_{01} \\ 
m_{10} & m_{11}%
\end{array}%
\right) ,\hat{K}_{1}=\left( 
\begin{array}{ll}
n_{00} & n_{01} \\ 
n_{10} & n_{11}%
\end{array}%
\right) ,  \label{kelements}
\end{equation}%
it follows that the coefficients $a_{nlm}(t)$ of the evolved state 

\begin{equation}
\left\vert {\varphi (t)}\right\rangle =\sum_{n,l,m=0,1}a_{nlm}(t)\left\vert
nlm\right\rangle _{ABE}  \label{estadot}
\end{equation}%
depend on the matrix elements $m_{ij}(t)$ and $n_{ij}(t),$ as well as on the
four constants $\alpha ,\beta ,\gamma $ and $\delta $ of the initial state %
\eqref{psi0}.

From the evolved state it is possible to study how the entanglement is distributed between 
$A$, $B$ and $E$. First, we analyze the amount of entanglement between the different pairs of subsystems during the evolution. The entanglement between pairs of qubits can be evaluated through the concurrence 
$C_{ij}$ \cite{wootters98}, where 
\begin{equation}
C_{ij}=\max \left\{ 0,\sqrt{\lambda _{1}}-\sqrt{\lambda _{2}}-\sqrt{%
\lambda _{3}}-\sqrt{\lambda _{4}}\right\},
\label{eq:concdef}
\end{equation}
 and the $\lambda _{i}$'s are the
eigenvalues (in decreasing order) of the matrix $\rho _{ij}(\sigma
_{y}\otimes \sigma _{y})\rho _{ij}^{\ast }(\sigma _{y}\otimes \sigma _{y})$, with $\rho_{ij} = \,$Tr$_{k}(\rho_{ijk})$.
Second, we analyze the possible emergence of genuine tripartite entanglement, meaning that the system is not separable in any bipartition. We show that some of the initial entanglement
can be converted into genuine tripartite entanglement. A possible measure of tripartite entanglement is the 3-tangle $\tau _{ijk}$, defined through the Coffman-Kundu-Wootters (CKW) relation \cite{ckw00} as:
\begin{equation}
\tau _{ijk}(t)=C_{i(jk)}^{2}(t)-C_{ij}^{2}(t)-C_{ik}^{2}(t),
\label{descomposicion}
\end{equation}
where the tangle $C^2_{i(jk)}=2(1-{\rm Tr}\rho_i^2)$ gives the amount of bipartite entanglement between $i$ and $jk$. This is an example of a monogamy relation between tripartite and bipartite entanglement. We note that monogamy relations of this sort have been investigated in Ref. \cite{Ma}.
  The 3-tangle $\tau _{ijk}$ is invariant under permutation of its indices, and if it takes a value different from $0$, then $C_{i(jk)}$ is also different from zero for \emph{all} $i,$ so that the system is not separable in any bipartition, or equivalently, the state has genuine
tripartite entanglement. Thus, $\tau _{ijk}\neq 0$ is a sufficient condition for identification of genuine
tripartite entanglement, though it is not necessary. Indeed, we can see from
Eq. (\ref{descomposicion}) that for $\tau _{ijk}=0,$ genuine tripartite entanglement
(i.e. non-biseparability) will be present whenever at least two qubit-qubit
tangles $C_{ij}^{2}(t)$ are non-zero \cite{durr00}. Therefore, 
the non-biseparable 3-qubit pure states are divided into two classes: those for
which $\tau _{ijk}\neq 0,$ and those for which $\tau _{ijk}=0$ and 
$C_{ij}\neq 0$ for two arbitrary pairs $ij.$ These classes coincide with the
two families of genuinely entangled states discussed in \cite{durr00},
namely the GHZ ($\tau _{ijk}\neq 0$) and the W ($\tau _{ijk}=0$) states.
Since the 3-tangle is nonzero for the GHZ-type family, $\tau _{ijk}$ can be
considered as a quantitative measure of genuine (GHZ-type) entanglement,
while for the W family there is no consensual measure of genuine
(W-type) entanglement. 
\par
For states of the form (\ref{estadot}), $\tau_{ijk}$ 
may be expressed in terms of the coefficients $a_{nlm}$ \cite{ckw00}, 
which are written in terms of the matrix elements of $\hat{K}_{0}$ and $\hat{K}_{1}$: 
\begin{equation}
\tau _{ABE}(t)=\mathcal{E}_{0}^{2}\left\vert f(\hat{K}_{0}(t),\hat{K}%
_{1}(t))\right\vert ,  \label{tau2}
\end{equation}%
where $f$ is the function 
\begin{gather}
f(\hat{K}_{0}(t),\hat{K}_{1}(t))=\left( m_{10}n_{01}-m_{00}n_{11}\right)
^{2}+  \label{efe} \\
+\left( m_{11}n_{00}-m_{01}n_{10}\right) ^{2}+  \notag \\
+2\left( m_{11}n_{01}-m_{01}n_{11}\right) \left(
m_{00}n_{10}-m_{10}n_{00}\right) -  \notag \\
-2\det K_{0}\det K_{1},  \notag
\end{gather}%
and $\mathcal{E}_{0}^{2}$ stands for the initial entanglement between $A$
and $B$: $\mathcal{E}_{0}^{2}=C_{AB}^{2}(0)=4\left\vert \alpha \delta
-\gamma \beta \right\vert ^{2}$.

The factorization of Eq. (\ref{tau2}) shows that the 3-tangle $\tau _{ABE}$ is
completely determined by the specific map and the initial bipartite entanglement,
and emerges as a redistribution --- induced by the local interaction of the system $B$ with
its environment --- of the initial bipartite entanglement $\mathcal{E}_{0}$
between $A$ and $B.$  Moreover, as a result of the $BE$
interaction, the bipartite entanglement $\mathcal{E}_{0}$ is not only
transformed into 3-tangle, but can also be distributed exclusively as
bipartite entanglement between couples of qubits, evolving to W-family
states, characterized by $\tau _{ABE}=0$. 
%%%%%%%%%%%%%%%%%%%%%%%%%%%%%%%%%%%%%%%%%%%%%%%
Let us now consider two concrete examples.
\subsection{Phase Damping Channel}
\label{sec:PDtheory}
The PD channel can be represented by the quantum map
\begin{subequations}
\label{eq:pd}
\begin{align}
\left\vert {0}\right\rangle _{B}\left\vert {0}\right\rangle _{E}&
\rightarrow \left\vert {0}\right\rangle _{B}\left\vert {0}\right\rangle _{E},
\\
\left\vert {1}\right\rangle _{B}\left\vert {0}\right\rangle _{E}&
\rightarrow \sqrt{1-p}\left\vert {1}\right\rangle _{B}\left\vert {0}%
\right\rangle _{E}+\sqrt{p}\left\vert {1}\right\rangle _{B}\left\vert {1}%
\right\rangle _{E}.
\end{align}%
\end{subequations}
Here $p$ is a parameter characterising the evolution, such that $p$ varies from $0$ to $1$.  To see the emergence of multipartite entanglement, consider the initial state:  
\begin{equation}
\label{phizero}
\left\vert {\varphi (0)}\right\rangle _{ABE}=\left[ \alpha \left\vert {11}%
\right\rangle +\delta \left\vert {00}\right\rangle \right] _{AB}\left\vert {0%
}\right\rangle _{E},
\end{equation}%
where $|\alpha|^2 + |\delta|^2 = 1 $. 
According to Eq. \eqref{eq:pd}, the tripartite system evolves to 
\begin{equation}
\left\vert {\varphi (p)}\right\rangle _{ABE}=\alpha \sqrt{1-p}\left\vert {110%
}\right\rangle +\alpha \sqrt{p}\left\vert {111}\right\rangle +\delta
\left\vert {000}\right\rangle.  \label{psideph}
\end{equation}%
We can see from Eq.(\ref{psideph}) that the system always evolves to a state belonging to the family of GHZ states \cite{acin01}. For this family of states, the 3-tangle $\tau _{ABE}$ of Eq. (\ref{tau2}) is a useful indicator of GHZ-type genuine tripartite entanglement. From Eqs. (\ref{tau2}) and (\ref{efe}),
with $m_{01} = m_{10} = n_{00} = n_{01} = n_{10} = 0$, $m_{00} = 1$, $m_{11} = \sqrt{1-p}$, and $n_{11} = \sqrt{p}$, one finds that the 3-tangle is given by: 
\begin{equation}
\tau _{ABE}=\mathcal{E}_{0}^{2}p.  \label{taudeph}
\end{equation}%
The maximally-entangled GHZ state has also a particular property: tracing out one of
the subsystems destroys any entanglement present. Therefore, for the state $%
\rho _{ij}=$Tr$_{k}$\thinspace $\rho _{ijk}$, we have $C_{ij}(\rho _{ij})=0.$
This is the case for the final state $\left\vert {\varphi (p=1)} \right\rangle _{ABE}$, as one can see 
from the expressions for the bipartite tangles as a function of $p$: 
\begin{eqnarray}
C_{AB}^{2}(p) &=&\mathcal{E}_{0}^{2}\!(1-p),  \label{tabledeph} \\
C_{BE}^{2}(p) &=&0,  \notag \\
C_{AE}^{2}(p) &=&0,  \notag
\end{eqnarray}%
where $C_{ij}$ was defined in Eq. (\ref{eq:concdef}).

%%%%%%%%%%%%%%%%%%%%%%%%%%%%%%%%%%%%%%%%%%
\subsection{Amplitude Damping Channel}
\label{sec:ADtheory}
Consider now the Amplitude Damping (AD) channel \cite{chuang00},  described by the quantum map
\begin{subequations}
\label{eq:ad}
\begin{align}
\left\vert {0}\right\rangle _{B}\left\vert {0}\right\rangle _{E}&
\rightarrow \left\vert {0}\right\rangle _{B}\left\vert {0}\right\rangle _{E},
\\
\left\vert {1}\right\rangle _{B}\left\vert {0}\right\rangle _{E}&
\rightarrow \sqrt{1-p}\left\vert {1}\right\rangle _{B}\left\vert {0}%
\right\rangle _{E}+\sqrt{p}\left\vert {0}\right\rangle _{B}\left\vert {1}%
\right\rangle _{E}.
\end{align}
\end{subequations}
%The AD channel \cite{chuang00}
% is equivalent to a two-level system decay, for instance.
 
When map \eqref{eq:ad} is applied to the initial
state 
\begin{equation}
\left| {\eta (0)}\right\rangle _{ABE}=\left[ \beta \left| {10}\right\rangle
+\gamma \left| {01}\right\rangle \right] _{AB}\left| {0}\right\rangle _{E},
\end{equation}
 it evolves to 
\begin{equation}
\left| {\eta (p)}\right\rangle _{ABE}=\gamma \sqrt{1-p}\left| {010}%
\right\rangle +\gamma \sqrt{p}\left| {001}\right\rangle +\beta \left| {100}%
\right\rangle .  \label{etaamp}
\end{equation}
State \eqref{etaamp} belongs to the W-family of states \cite{acin01}. 
As mentioned before, the W-family is characterized by
null 3-tangle $\tau _{ABE}$ and nonzero tangles $C_{ij}^{2}$.
Tracing out one of the subsystems leaves the other two entangled. 
The tangles for the pure state (\ref{etaamp}) as a function of $p$ are given by
\begin{eqnarray}
C_{AB}^{2}(p) &=&\!\mathcal{E}_{0}^{2}(1-p),  \label{tableamplitude} \\
C_{BE}^{2}(p) &=&4|\gamma|^4 p(1-p),  \notag \\
C_{AE}^{2}(p) &=&\mathcal{E}_{0}^{2}p.  \notag
\end{eqnarray}
Thus, every $C_{ij}^{2}$ is different from zero in the interval $\left( 0,1\right) ,$ 
and it follows that genuine W-type entanglement should be present.

%%%%%%%%%%%%%%%%%%%%%%%%%%%%%%%%%%%%%%%%%%%%%%%%%%%%%%%%%%
\section{Experiment}
In this section, we experimentally investigate the entanglement redistribution  discussed in the last section.  A number of experiments have used photonic degrees of freedom to investigate the dynamics of open quantum systems \cite{kwiat00,aiello07,almeida07,salles08,xu10,farias12,farias09,farias12b,jeong13}.  Usually, the polarization degree of freedom represents the qubit system and the spatial or spectral degrees of freedom play the role of the environment.  In this way, birefringent material or interferometers can be used to implement system--environment interactions. 
\label{sec:experiment}
%%%%%%%%%%%%%%%%%%%%%%%%%%%%%%%%%%%%%%%%%%%%%%%%%%%%%%%%%%
\subsection{Experimental Setup}
%%%%%%%%%%%%%%%%%%%%%%%%%%%%%%%%%%%%%%%%%%%%%%%%%%%%%%%%%%
In our experiment, we produce a pair of photons entangled in polarization and implement
quantum channels for one of the photons using an interferometer. The
environment in this case may be considered as a two-level system, its degrees of freedom being
the two different propagation modes of the interferometer. We
are able to measure both populations and coherences of the environment
using a second interferometer.  In this way, we have access to the tripartite 
system $ABE$, and complete three-qubit quantum state tomography can be performed. 
\par
\begin{figure}[tbp]
\centering 

  \includegraphics[width=8cm]{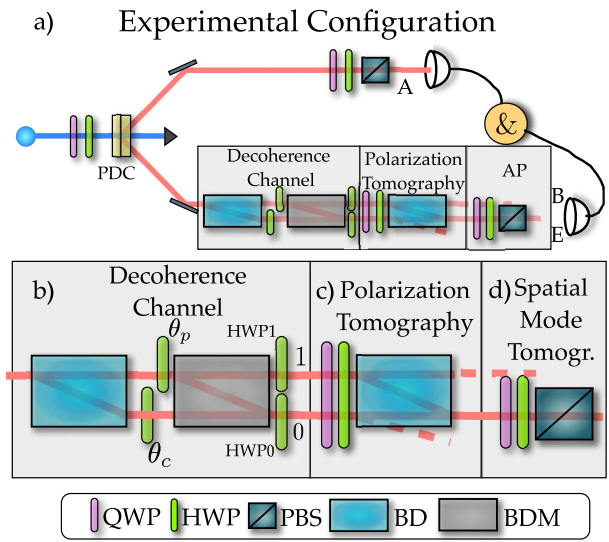}

\caption{Experimental Setup.}
\label{fig:setup}
\end{figure}

The experimental setup is shown in Fig. \ref{fig:setup}. A $\lambda_p=325$ nm wavelength He-Cd (Helium-Cadmium) laser pumps two non-linear crystals producing polarization-entangled photon pairs,
both with central wavelengths at $650 $nm \cite{kwiat99}. The polarization state of the pair can be written as:
\beq
\ket{\psi}_{AB}= \alpha_1 \ket{HH}_{AB} + \alpha_2 \ket{VV}_{AB},
\eeq
where $H$ and $V$ stand for horizontal and vertical polarization directions. The coefficients 
$\alpha_i$ ($ i=1,2 $) can be controlled through the polarization of the pump laser 
(see Fig.\ref{fig:setup}), because each polarization component ($H$ or $V$) pumps one of the
crystals. In the following,  we let $H$ ($V$) represent the 0 (1) state in the computational basis.  
Photon $A$ goes directly to polarization analysis, which is performed using a QWP (quarter waveplate), 
a HWP (half waveplate), a PBS (polarizing beam splitter), and a single photon counting module (SPCM). 
Photon B is sent through two nested interferometers. The first one, shown in Fig \ref{fig:setup}b), is  responsible for the implementation of the quantum maps given by Eq. (\ref{Map}). This is done by coupling 
the spatial modes of the interferometer with the polarization components of the photons.
The output modes of the interferometer are labeled $0$ and $1$.  
Similar coupling between spatial modes modes and polarization was obtained in previous
experiments, using a Sagnac-like interferometer \cite{almeida07, salles08, farias09}. 
\par
Let us discuss the first interferometer, shown in Fig. \ref{fig:setup} b), in more detail. The incoming beam passes through a birefringent calcite beam displacer (BD). The input beam is split so that the $H$ polarized component is refracted and the $V$ polarized component is transmitted. After the BD, the $H$ and $V$ components are displaced from one another and propagate in parallel directions, each one through a half waveplate. The $H$ component passes through  HWP($\theta_c$) and the $V$ component through HWP($\theta_p$). In this way the polarization
components can be rotated independently. In order to implement the quantum channels, it is necessary to set 
$\theta_c =0$ and $\theta_p$ varying  within the interval $0\leq \theta _{p}\leq \pi /4$. HWP($\theta_c$)
only compensates the path difference due to the presence of HWP($\theta _{p}$) in the other path. 

The beams are recombined in the BDM(see Fig. \ref{fig:setup}), which transmits the $H$ polarized photons and deflects the $V$ polarized photons. The BDM is realized by  placing a BD between two HWP's(not shown), which convert $H$ into $V$ and vice versa, before and after the BD. 
This is necessary for the recombination in the BDM, otherwise the beams would split 
instead of recombining. When $\theta_p = 0$, the polarization components split in the first BD, are
coherently recombined in the BDM, and the input polarization state is recovered at the otput 
of BDM. When $\theta_p \neq 0$, one polarization component can be increased while the other is decreased, 
according to the setting of $\theta_p$.

This transformation is described by the quantum map given by the AD channel \eqref{eq:ad},
where the states $|0\rangle_E$ and $|1\rangle_E$ correspond to the spatial modes 0 and 1 in Fig. 2 b), 
and $p=\sin^2 (2\theta _{p})$ is a parameter related to the evolution of the interaction
between subsystems $B$ and $E$. The correspondence between this parameter and the time in
a time-dependent interaction is given by \thinspace $p(t=0)=0$ and $p(t=\infty )=1$. 

If one wants to implement the Phase Damping (PD) channel shown in Eqs. \eqref{eq:pd}, one should set HWP1 to 
$\pi /4$, so that the polarization component $H$ propagating along mode 1 at the output of the first interferometer is completely converted into $V$. The role of HWP0, which is fixed at $0{{}^o}$, 
is to balance the optical paths inside the second interferometer. At this point, one can see clearly
the difference in the implementation of the AD and PD channels. In both cases, part of the $V$ component
of the input state is changed into $H$ inside the first interferometer and sent to another spatial mode.
If one detects modes $0$ and $1$ without distinguishing one from the other (tracing out the spatial modes),
then the AD channel is implemented. However, if we use HWP1 to turn the polarization of mode $1$ back to
$V$ and detect tracing out the spatial modes, the PD channel is implemented.

%%%%%%%%%%%%%%%%%%%%%%%
%%%%%%%%%%%%%%%%%%%%%%%

The second interferometer and the Spatial Mode Tomography (SMT)  shown in Figs. \ref{fig:setup} c) and \ref{fig:setup} d), are used to analyze the polarization and spatial qubits simultaneously. 
The polarization analysis is made using the QWP, the HWP, and the third BD in Fig. \ref{fig:setup} c) which acts as a polarizer (only the lower output is used). 
At the same time, we can see that the lower output of the third BD is a combination of two modes: the $H$-polarized component coming from spatial mode $1$ of subsystem $E$, and the $V$-polarized component coming 
from spatial mode $0$ of subsystem $E$. In this way we can measure the populations and coherence between modes $0$ and $1$ of subsystem $E$, because this information is swapped to the polarization degree of freedom. Therefore, for each setting of the QWPs and HWPs in Figs. \ref{fig:setup} c) and d), a given measurement on
the spatial modes of subsystem $E$ and polarization modes of subsystem $B$ is made, and full tomography can
be performed. 

After the interferometers and the SMT, the photons are detected with a single-photon counting module (SPCM), and coincidence counts are registered. The dynamics of the entanglement is investigated for both channels. The evolution of
the interaction is controlled through the waveplate $HWP(\theta_p)$, and reconstruction of the complete tripartite states is performed. This requires 64 different settings of the three sets of (QWP,HWP) waveplates. This new configuration represents an advance in comparison with that of Refs. \cite{almeida07,salles08,farias09}, as it is more stable and allows full tomographic access to polarization and spatial modes. 

%%%%%%%%%%%%%%%%%%%%%%%%%%%%%%%%%%%%%%%%%%%%%%%%%%%%%%%%%%%%%%%%%%%%%%%%%%%%%
\section{Experimental Results}
\label{results1}
%%%%%%%%%%%%%%%%%%%%%%%%%%%%%%%%%%%%%%%%%%%%%%%%%%%%%%%%%%%%%%%%%%%%%%%%%%%%%

%%%%%%%%%%%%%%%%%%%%%%%
\begin{figure}
\centering 

  \includegraphics[bb=31 2 815 581,width=2.99in,height=2.39in,keepaspectratio]{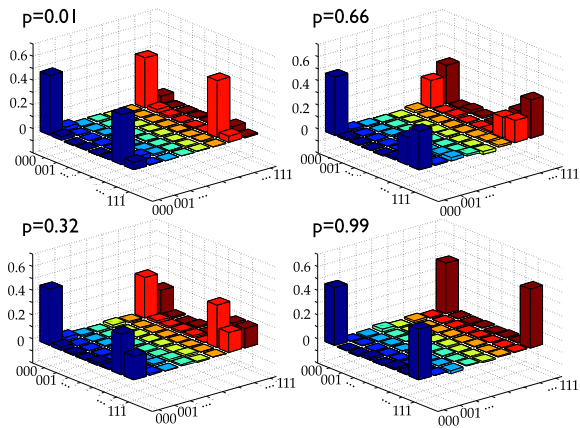}

\caption{(Color online). Real part of the density matrices reconstructed by quantum state tomography for different values of $p$. $B$ and $E$ evolve according to the Phase Damping channel.}
\label{fig:densidaddeph}
\end{figure}
%%%%%%%%%%%%%%%%%%%%%%%
We first implemented the PD channel, starting from an initial state close to that of Eq. \eqref{phizero}.    We vary the control parameter $p$ and for each setting we perform full quantum state tomography, including systems $A$, $B$, and $E$. The real part of the experimentally reconstructed density matrices can be seen in  Fig. \ref{fig:densidaddeph}. The imaginary parts are close to zero and are not shown. The initial state is prepared so that $|\alpha|^2 \simeq |\delta|^2 \simeq 1/2$, using
the wave plates in the pump beam (see Fig.\ref{fig:setup}). Qualitatively, we observe that as $p$ increases, the populations $|111\rangle \langle 111|$ and coherences $|111\rangle \langle 000|$, $|000\rangle \langle 111|$ increase, while other
contributions ($|110\rangle \langle 110|$ and $|110\rangle \langle 000|$) decrease.  This leads to a GHZ state when $p$ approaches $1$. 
%%%%%%%%%%%%%%%%%%%%%%%
\begin{figure}[tbp]
\centering 

  \includegraphics[bb=50 7 801 590,width=3.06in,height=2.45in,keepaspectratio]{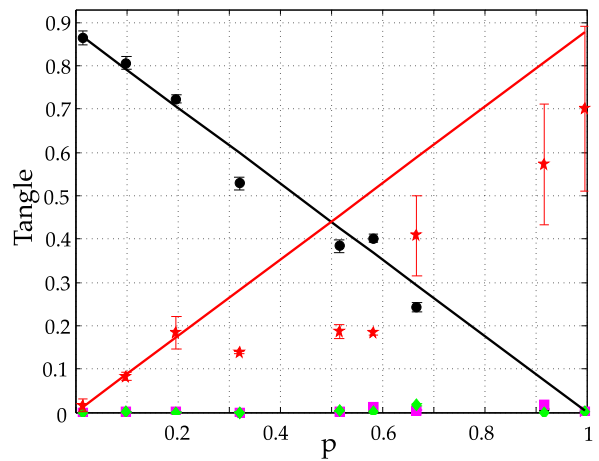}

\caption{(Color online) Evolution of the tangles as a function of $p$, for the phase damping
channel. While $C_{AB}^{2}(p)$ (black circles) decays linearly, the 3-tangle $\tau(p)$ (red stars)
grows with $p$. The bipartite entanglement between qubits $A$ and $B$ and the environment $E$, 
given by $C_{AE}^{2}(p)$ (green diamonds)and $C_{BE}^{2}(p)$ (magenta squares) respectively, 
is zero along the whole evolution. }
\label{fig:dephcompleto}
\end{figure}
%%%%%%%%%%%%%%%%%%%%%%%
From the reconstructed density matrices we calculate the concurrence and tangle for all bipartitions.  The concurrence $C_{ij}(p)$ between qubits $i$ and $j$ is calculated using the definition of 
Eq. \eqref{eq:concdef}. 

We also calculate the 3-tangle from the CKW relation in Eq. \eqref{descomposicion}.  To do so, it was necessary to calculate the concurrence $C_{i(jk)}(p)$, taking advantage of the fact that the measured states can be considered quasi-pure, with average fidelities around 
$0.90$ with respect to the pure states of Eq. (\ref{psideph}).  
The concurrence for quasi-pure states of any dimension that present a predominant
eigenvalue in the spectral decomposition can be approximated by 
$C_{i(jk)} \approx \mathrm{max}(0,\varsigma _{1}-\sum_{i>1}\varsigma_{i})$, 
where the $\varsigma_i$'s are the positive eigenvalues of a positive Hermitian matrix that is defined in terms of the eigenvectors and eigenvalues of
the density matrix, as described in detail in Ref. \cite{mintert05b}. Substituting this expression into Eq.~(\ref{descomposicion}) gives the three-tangle for the quasi-pure state.
 \par

In Fig. \ref{fig:dephcompleto} we plot the tangles obtained from the
reconstructed states (points) and the theoretical prediction (curves) as explained above. The bipartite 
entanglement of system $AB$ (black circles) is in good agreement with theory given by Eq. (\ref{tabledeph}) (black line). 
Good agreement between theory and experiment is also obtained for the predicted null 
tangles $C_{BE}^{2}$ (magenta squares) and  $C_{AE}^2$ (green diamonds). However,
the tripartite entanglement $\tau_{ABE}$ (red stars) presents significant deviation
from the theory (red line), even though they qualitatively show the same tendency. 
This discrepancy is most likely due to the fact that the theory is valid for pure states and the calculation of  $\tau_{ABE}$ is made with 
the CKW relation \cite{ckw00} given by Eq. (\ref{descomposicion}) under the assumption that the states are quasi-pure. We conclude that the 3-tangle is very sensitive
to noise, assuming that the quasi-pure approximation is appropriate in this case. 
The error bars in all figures were produced from Monte-Carlo simulations of experimental results obeying the 
same count statistics.  
In Fig.  \ref{fig:dephcompleto} we see from the error bars that the signal
to noise ratio decreases when $p$ increases. We also observe that points around $p=0.5$
tend to have lower values of $\tau_{ABE}$. This is probably due to the fact that when $p=0.5$, 
the photonic modes of subsystems $B$ and $E$ are distributed equally between the different paths of the interferometer, and are thus more sensitive to phase fluctuations and imperfect mode matching.
\par
We also investigate experimentally the distribution of entanglement for
the case of the AD channel. 
%%%%%%%%%%%%%%%%%%%%%%%
\begin{figure}[tbp]
\centering 

  \includegraphics[bb=32 7 809 580,width=3.06in,height=2.14in,keepaspectratio]{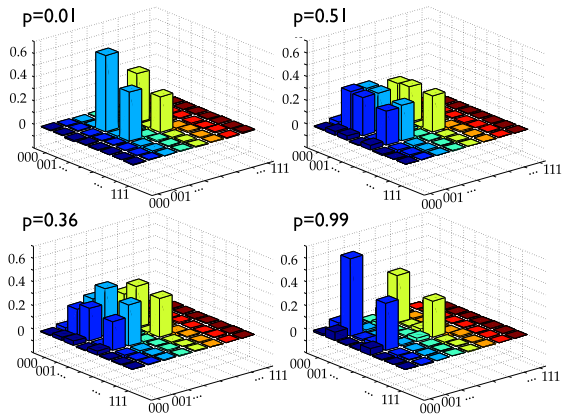}

\caption{(Color online). Real part of the density matrices obtained from 
tomographic reconstruction for different values of $p$, when $B$ is under the action of an Amplitude 
Damping channel.}
\label{fig:densidadampl}
\end{figure}
%%%%%%%%%%%%%%%%%%%%%%%
The real part of the measured density matrices for this quantum channel is shown in Fig. \ref{fig:densidadampl}.   The imaginary parts are close to zero in all cases, and are not shown. 
In this case we can see that the populations $|001\rangle \langle 001|$, as well as the coherences 
$|001\rangle \langle 100|$, increase with $p$, while $|010\rangle \langle 010|$ and $|010\rangle \langle 100|$ decrease. When $p=0.5$ the state is approximately a maximally entangled $W$ state. 
Moreover, we see that the evolution results in an entanglement swapping between subsystems $AB$ and $BE$. 
At $p=0$, all the entanglement is between systems $A$ and $B$, and
at $p=1$ all the entanglement is transferred to subsystem $A$ and the environment $E$. 

%%%%%%%%%%%%%%%%%%%%%%%
\begin{figure}[tbp]
\centering 

  \includegraphics[bb=48 5 801 583,width=2.99in,height=2.39in,keepaspectratio]{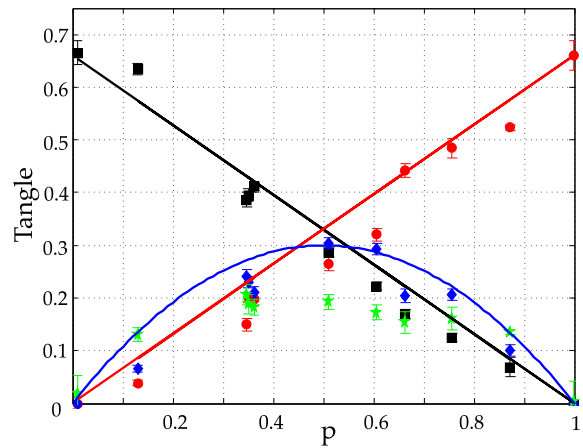}

\caption{(Color online). Bipartite tangles for the Amplitude Damping
channel. $C_{AB}^{2}(p)$ (black squares) and $C_{AE}^{2}(p)$ (red circles).
$C_{AB}^{2}(p)$ decays and $C_{AE}^{2}(p)$ grows linearly with $p$. 
The entanglement between $B$ and $E$, $C_{BE}^{2}(p)$ (blue diamonds), evolves
quadratically. The 3-tangle (green stars) is nonzero, contrary to what is expected
for a pure state.}
\label{fig:amplcompleto}
\end{figure}
%%%%%%%%%%%%%%%%%%%%%%%
From the density matrices we calculate the tangles for the case of the AD channel, which we show in 
Fig. \ref{fig:amplcompleto}. 
The solid lines correspond to the theoretical predictions of Eq. \eqref{tableamplitude} for each case. 
We notice that the experimental points are in good agreement with theory for the bipartite entanglement. 
We observe a nonzero and non-negligible 3-tangle along the evolution, even though
it should be zero if the initial state were perfectly pure. 
This unexpected appearance of a non-null 3-tangle is a manifestation of the high sensitivity 
of $\tau$ to the  impurities of the states. This conclusion comes from the observation 
that $\tau$ is null for the family of states in Eq. (16), and even though the measured
states have fidelities as high as 0.9 with respect to those 
states, $\tau$ is significantly different from zero for them.

%This genuine entanglement, measured by $\tau_{ABE}$ within the quasi-pure 
%approximation, has a common characteristic with the same quantity in the context
%of the PD channel: $\tau_{ABE} $ is highly sensitive to the impurity of the states. 
%Also in this case, the measured states have all fidelities as high as $0.9$ with respect 
%to the pure states of the Eq. (\ref{etaamp}). However, they can present significantly
%different tripartite entanglement.

In this case it is especially interesting that noise, or the impurity of the states, can induce 
the appearance of $\tau\neq 0$. This means that the uncontrolled noise along the evolution couples 
states of the $W$-class to the $GHZ$-class. According to Ref. \cite{durr00} , this cannot be 
done by Stochastic Local Operations and Classical Communication (SLOCC). 
Therefore, we are led to conclude that some operation that is not independent for each degree of
freedom of the same photon is affecting our system. 
For instance, mode mismatching and phase fluctuations in the 
nested interferometers, may affect simultaneously polarization $B$ and spatial $E$ qubits.
We believe this is the most probable reason for the appearance of correlations coming from
the noise.

%tripartite entanglement. In this example of the AD channel, the existence of
%non-zero 3-tangle indicates that the noise couples the initial bipartite entanglement to GHZ-type 
%of tripartite entanglement. This is surprising, since the experimental factors causing this noise 
%are related to degrees of freedom which are not considered in the state reconstruction process.
%For instance, mode mismatching in the interferometers and phase fluctuations are the most probable 
%reasons for the appearance of mixed components. On the one hand, we have seen that the PD channel
%(which is essentially phase noise) leads to GHZ-type of entanglement, but only when 
%the environment is a part of the reconstructed entangled state. On the other hand, when we measure 
%the coherences of the state using the {\em imperfect} interferometers, we  ``trace out" the 
%non-overlapping modes. That is, the environment is not included in the reconstructed state as in the
%case of the PD channel. Therefore, the most probable cause of this ``unexpected´´ tripartite 
%entanglement is that the noise is acting as global environment for the two degrees of freedom
%of the photon. 

%%%%%%%%%%%%%%%%%%%%%%%%%%%%%%%%%%%%%%%%%%%%%%%%%%%%%%%%%%%%%%%%%%%%%%%%%%%%%
\subsection{Decomposition in Pure States}
\label{sec:decomp}	
%%%%%%%%%%%%%%%%%%%%%%%%%%%%%%%%%%%%%%%%%%%%%%%%%%%%%%%%%%%%%%%%%%%%%%%%%%%%%

To observe the theoretical predictions presented in section \ref{sec:theory}, we would ideally like to prepare initial pure states of the form given by Eq. (\ref{psi0}),
 apply the PD and the AD channels to a subsystem, and measure the evolved state including
the environment. In practice, we observe that the overall purity of the measured states 
is not unity, though always higher than $0.8$. The undesired mixture is due to technical problems like misalignment of the interferometers, small intensity fluctuations of the pump laser, and other issues.

%%%%%%%%%%%%%%%%%%%%%%%
\begin{figure}[tbp]
\centering 

  \includegraphics[bb=5 36 585 814,width=2.99in,height=3.74in,keepaspectratio]{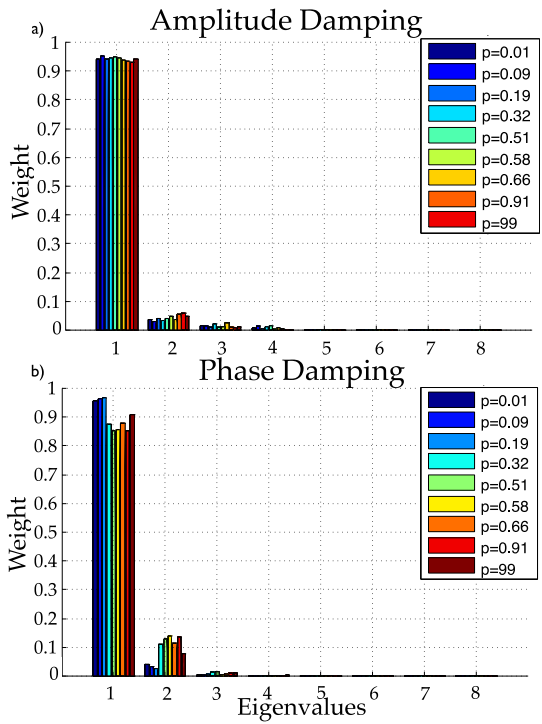}

\caption{(Color online). Eigenvalues of the measured density matrices for the 
two channels implemented. In all states measured there is always one
eigenvalue higher than $0.85$, so that we can consider that the states are quasi pure.}
\label{fig:autovalores}
\end{figure}
%%%%%%%%%%%%%%%%%%%%%%%
The density matrices shown in Figs. \ref{fig:densidaddeph} and \ref{fig:densidadampl} show qualitatively that multipartite GHZ-type and W-type entanglement results from the decoherence process.  However, Figs. \ref{fig:dephcompleto} and \ref{fig:amplcompleto} show that there is some deviation from theory, presumably due to the mixedness of the experimental quantum states.  
\par
To further investigate the role of mixedness, we first diagonalize the experimentally obtained density matrices. Fig. \ref{fig:autovalores}a) shows the spectral decomposition

\begin{equation}
\rho = \sum \mu_i \ket{\phi_i}\bra{\phi_i},
\label{spec:decompose}
\end{equation}
of the measured states when the AD channel is implemented. We notice that for all values of 
$p$ there is a dominant eigenvalue which is always greater than $0.92$.
Fig. \ref{fig:autovalores}b) shows the spectral decomposition
of the measured states when the PD channel is implemented. We also notice that for all values 
of  $p$ there is a dominant eigenvalue which is always greater than $0.85$.
Therefore, we conclude that  the unpredicted effects coming from the impurity are due to a
small contribution of non-dominant eigenvectors.  However, as Figs. \ref{fig:dephcompleto} and \ref{fig:amplcompleto} show, some of these effects are significant.
\par
The non purity of the states affects the tripartite much more than the bipartite entanglement. 
In the case of the PD channel, the values of the 3-tangle tend to
be smaller for $p \leq 0.3$ where the contribution of other spectral components is larger, as we can see
in Fig. \ref{fig:autovalores}b).
For the AD channel, the relative contribution of other spectral components is roughly
constant when $p$ is varied, as can be seen from Fig. \ref{fig:autovalores} a).  However, as we saw in Fig.  \ref{fig:amplcompleto},  the 3-tangle (which
is zero for pure states) is greater in the region around $p=0.5$. Even though the reason why 
the impurity induces GHZ-type entanglement is unclear, it is rather intuitive that the region around $p=0.5$ is more critical due to the near equiprobable distribution of photon B in the two spatial modes of the environment E.

%We notice that the PD channel leads to states presenting maximum eigenvalues which are lower 
%han those of the states obtained from the AD channel. 
%he smaller eigenvalue for the PD channel, corresponding to $p=0.91$,
%s $0.85\pm 0.02$. This is caused by lack of purity of the measured
%states that populate the spectral components $2$ and $3$ of Fig. (\ref%
%{fig:autovalores}b). As we see in the previous section,  this noise in the measurements
%deviates a lot the experimental results for the 3-tangle with respect to the theoretical
%predictions.  For the case of the PD for instance,  we can see from Fig. \ref{fig:autovalores}b)
%that the mixedness is bigger for values of $p$ between $0.3$ and $0.95$. Just for this values, 
%%we can see from Fig. \ref{fig:dephcompleto}, that the separation 
%between the theoretical predictions and the measured states is the greater. For the cases of the AD,
%there are small components of noise in the measurements, this can be verified observing the
%small value of components $2$ and $3$ of the Fig \ref{fig:autovalores}a). This small components of mixedness 
%generate generate values of 3-tangle different from zero from this channel. 
%This will become clearer in the following section.

%%%%%%%%%%%%%%%%%%%%%%%%%%%%%%%%%%%%%%%%%%%%%%%%%%%%%%%%%%%%%%%%%%%%%%%%%%%%%
The existence of a predominant eigenvalue for all the states
justifies the quasi-pure approximation used in \cite{farias12b}. To determine whether the impurity of the states is indeed responsible 
for the deviation of the experimental points from the theory for pure states,  in what follows we will analyze the data by considering only the pure state 
corresponding to the dominant eigenvector.
\subsection{Entanglement Dynamics for Pure States}
%%%%%%%%%%%%%%%%%%%%%%%%%%%%%%%%%%%%%%%%%%%%%%%%%%%%%%%%%%%%%%%%%%%%%%%%%%%%%
%%%%%%%%%%%%%%%%%%%%%%%
\begin{figure}[tbp]
\centering 
 
  \includegraphics[bb=54 6 796 588,width=3.06in,height=2.45in,keepaspectratio]{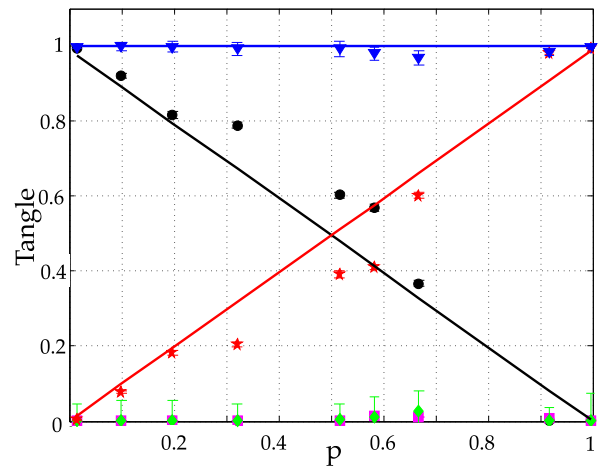}

\caption{(Color online). Bipartite and multipartite entanglement as a
function of $p$ for the PD channel, considering only the dominant eigenvalue. 
The 3-tangle $\protect\tau_{ABE}$ 
(red stars) increases linearly as the bipartite entanglement $C_{AB}^{2}$ 
decreases (black circles). The entanglement in bipartition $BE$, $C_{BE}^{2}$ 
(green diamonds) and $AE$, $C_{AE}^{2}$ (magenta squares) are nearly zero during
the entire evolution. The solid lines are the theoretical predictions given by
Eq. \eqref{tabledeph}. The blue down triangles show the experimental values of the
invariant given by Eq. \eqref{dep}, and the blue solid line gives the corresponding 
theoretical values. }
\label{fig:tanglesdeph}
\end{figure}
%%%%%%%%%%%%%%%%%%%%%%%
From now on, we analyze the entanglement dynamics for the largest component of the spectral 
decomposition, that is, the eigenstate $\ket{\phi_m}$ associated to the eigenvalue $\mu_m=\rm{max}[\mu_i]$.
We begin analyzing the data corresponding to the PD channel. 
Fig. \ref{fig:tanglesdeph} shows the dynamics of entanglement for this channel. 
The experimental results for the tangles as a function of $p$ are shown: 
$C_{AB}^{2}$ (black circles), $C_{BE}^{2}$ (green diamonds), and $C_{AE}^{2}$
(magenta squares). The solid lines are theoretical predictions given by
Eq. (\ref{tabledeph}).
The 3-tangle $\tau _{ABE}$ is also shown. 
The theoretical value given by Eq. (\ref{taudeph}), corresponds to the
red solid line, and the experimental points (red stars) are obtained
from the expression 
\begin{equation}
\tau _{ABE}=4\det \rho _{A}-C_{AB}^{2}(\rho _{AB})-C_{AE}^{2}(\rho _{AE}).
\label{tauexp}
\end{equation}
This result is obtained from Eq. (\ref{descomposicion}), when one considers 
that for pure states  $C_{A(BE)}^{2}$ can be rewritten as 
$2\left( 1-\text{Tr}\rho _{A}^{2}\right) =4\det \rho _{A}$. 

While the data points are clearly closer to the theoretical prediction, we still notice a discrepancy  for 
$\tau_{ABE}$. The reason for this is related to the fact that the experimental points come 
from the most significant eigenstate in the spectral decomposition of each reconstructed density 
matrix. The more the reconstructed state is mixed, more the approximate pure state deviates
from the theoretical prediction. As seen in Fig. \ref{fig:autovalores}b), the
states with lower eigenvalues near $p \simeq 0.5$ (approximately  $0.85)$ correspond exactly to those points in 
Fig. \ref{fig:tanglesdeph} having more significant discrepancy from the theoretical prediction.  As mentioned above, when $p=0.5$, the photon is equally spread between the two arms of the interferometer, rendering phase fluctuations and mode-matching errors more significant.    
As we can see, the GHZ-type of entanglement, measured by $\tau _{ABE,}$ emerges as soon
as the interaction between system $B$ and its environment is switched on.
It increases linearly with $p$ until a GHZ state is reached. Moreover, in this analysis, 
it is evident that the increase of tripartite entanglement occurs at the expense of the initial bipartite
entanglement between systems $A$ and $B$. This follows from the expression 
\begin{equation}
\mathcal{E}_{0}^{2}=C_{AB}^{2}(p)\!+\tau _{ABE}(p),  \label{dep}
\end{equation}%
obtained from Eq. (\ref{taudeph}) and the first line of Eq. (\ref{tabledeph}). 
This shows that $C_{AB}^{2}(p)\!+\tau_{ABE}(p)$ is an invariant along the evolution, a result that generalizes
the one found in \cite{farias12}. This invariant is plotted the Fig. \ref{fig:tanglesdeph}
(blue line), and shows that the quantity $\mathcal{E}_{0}^{2}=C_{AB}^{2}(0)=0.99\pm 0.001$, 
though constant, changes its physical meaning depending on the value of $p$: it is totally bipartite
entanglement at $p=0$, and is completely transformed into tripartite
entanglement at $p=1$, when \emph{all }the qubit-qubit (bipartite)
entanglement vanishes. At intermediate stages $\mathcal{E}_{0}^{2}$ is
distributed in bipartite and GHZ-type of entanglement.

%%%%%%%%%%%%%%%%%%%%%%%
\begin{figure}[tbp]
\centering 

  \includegraphics[bb=37 6 801 594,width=3.06in,height=2.45in,keepaspectratio]{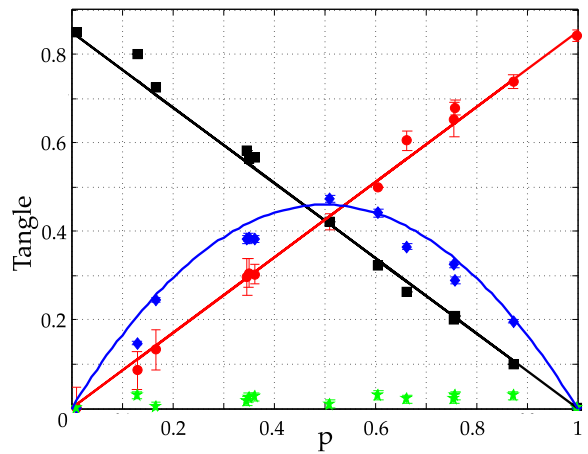}

\caption{(Color online). Bipartite tangles for the AD
channel and 3-tangle, considering only the dominant eigenvalue. 
$C_{AB}^{2}(p)$ (black squares) decays and $C_{AE}^{2}(p)$ (red circles)
grows linearly with $p$. The entanglement between
qubit $B$ and the environment $E$, $C_{BE}^{2}(p)$ (blue diamonds), evolves
quadratically. The 3-tangle (green stars) is nearly zero during the evolution.}
\label{fig:conc_amp}
\end{figure}
%%%%%%%%%%%%%%%%%%%%%%%
For the case of the AD channel, the evolution of the tangles is shown in Fig. \ref{fig:conc_amp}.
$C_{AB}^{2}(p)$ (black squares) decays, and $C_{AE}^{2}(p)$ (red
circles) increases linearly with $p$, whereas the entanglement between system 
$B$ and its environment, $C_{BE}^{2}(p)$ (blue diamonds), evolves quadratically. 
The theoretical predictions of Eq. (\ref{tableamplitude}) are the solid 
lines. We can see that there is a good agreement between theory and experiment in this analysis. 
For the 3-tangle, the theory predicts $\tau _{ABE}=0$ for all $p$, which is also in 
agreement with measurements in this approximation (green stars). 
This result emphasizes the idea that the mixed component of the states
generates the GHZ-type of entanglement discussed in the end of section \ref{results1}.
The plots show that the AD channel entangles $E$ and $B$ (with the entanglement vanishing 
only in the limits \thinspace $p=0,1$), and transfers the initial entanglement $\mathcal{E}_{0}$
from the pair $AB$ to the pair $AE$, characterizing the swapping process mentioned
in section \ref{results1}. This transfer is linear in $p$. 

%is a reflect of the
%invariant%
%\begin{equation}
%\mathcal{E}_{0}^{2}=C_{AB}^{2}(p)\!+C_{AE}^{2}(p)\!,  \label{invAD}
%\end{equation}%
%which is the equation analogous to Eq. (\ref{dep}) for the AD case. It is
%interesting to observe that the genuine entanglement for the W state does
%not intervene in the invariant.

\begin{figure}[tbp]
\centering 

  \includegraphics[bb=50 6 802 584,width=3.14in,height=2.51in,keepaspectratio]{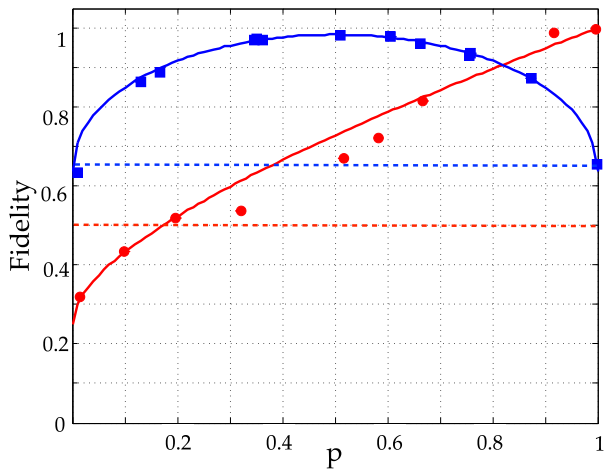}

\caption{(Color online) Fidelity with respect to multipartite entangled states. 
Red circles show the fidelity with respect to GHZ state in the case of the PD channel.
Blue squares show the fidelity with respect to W state in the case of the AD
channel. The solid lines correspond to theoretical evolutions of the initial
pure state. We can see that for the evolution of both channels, the emerging
states have genuine entanglement using this fidelity as witness 
\protect\cite{acin01}. The dashed lines represent the threshold for each one
of the witnesses.}
\label{fig:fidelidadespuras}
\end{figure}

As remarked in Ref. \cite{durr00}, the 3-tangle defined in Eq. (\ref{descomposicion}) 
is not sensitive to genuine tripartite entanglement for states
of the W family. Thus, the emergence of genuine entanglement in these cases
must be detected using other quantities or witnesses. Here we employ
the fidelity $F_{W}=\langle W|\rho |W\rangle $, which has been recognized
as a good witness to detect W-type entanglement \cite{acin01}. In
this case, the criterion states that a 3-qubit state $\rho $ has W-type
genuine entanglement whenever $F_{W} \geq 2/3$. 
Fig. \ref{fig:fidelidadespuras} shows that for $p\in \left( 0,1\right) $
the experimental fidelity $F_{W}$ (blue squares) is greater than $2/3$.
This confirms that the measured state has W-type entanglement.
%%%%%%%%%%%%%%%%%%%%%%%
%%%%%%%%%%%%%%%%%%%%%%%
As expected, $F_{W}$ reaches a maximum value at $p=0.5,$ when the state (\ref{etaamp}) 
ideally becomes a $W$ state. The experimental points are in very good agreement
with the theoretical predictions (blue line). 

For detecting genuine entanglement in states belonging to the GHZ family, we use 
$F_{GHZ}=\langle GHZ|\rho |GHZ\rangle $. In this case it has been shown that
a 3-qubit state $\rho $ is non biseparable whenever $F_{GHZ}>1/2$ \cite{acin01}, 
but the GHZ-type of 
entanglement is assured only if $F_{GHZ}>3/4.$ The experimental results for $F_{GHZ}$ 
are also shown in Fig. \ref{fig:fidelidadespuras} (red circles). 
We observe that for $p>0.2$ the measured states are non-biseparable. However, we cannot assure that the genuine entanglement
is of the GHZ type until $p=0.66,$ where $F_{GHZ}$ is larger than $3/4$. Comparison of the results obtained for $F_{GHZ}$ with those obtained for $\tau _{ABE},$ shows that this latter is a better indicator of 
GHZ-type of genuine entanglement, since $\tau _{ABE}$ is different from zero
already for $p>0.$ The discrepancy between the experimental points and the
theoretical curve (red solid line) is due to the mixed components of the measured states,
in the same way as in Fig. \ref{fig:tanglesdeph}. 

%%%%%%%%%%%%%%%%%%%%%%%%%%%%%%%%%%%%%%%%%%%%%%%%%%%
\subsection{Evolution of the quantum discord}
%%%%%%%%%%%%%%%%%%%%%%%%%%%%%%%%%%%%%%%%%%%%%%%

Over the last decade, it was demonstrated theoretically and experimentally that the QD could be a resource to improve some tasks \cite{datta05, dakic12} related to information processing. Furthermore, it was shown that
its dynamics may present non-analytic points \cite{maziero09}. This effect was demonstrated experimentally in an optical setup \cite{xu10}, and a physical interpretation of these abrupt changes was related to the quantum measurement problem in Ref. \cite{cornelio12}, where it was shown that the so called pointer basis may
emerge much before the decoherence process is significant. These results support the
idea that the analysis of the dynamics of the QD can shine new light on different kinds of problems
in physics and applications.

In the following, we investigate the dynamics of genuine (tripartite) QD (GQD) and 
total QD (TQD) theoretically and experimentally for the PD and AD channels. 
For this analysis, we utilize the tools introduced in
\cite{giorgi11}. The expressions 
for GQD and TQD are obtained by generalizing the usual definition of 
QD  \cite{ollivier02} to the tripartite case. This generalization presents 
the advantage
of quantifying quantum correlations in more general systems, even 
for mixed states. First, the total information
in a tripartite system $ABC$ is 
obtained and expressed as
\beq
T(\rho)=S(\rho_A)+S(\rho_B)+S(\rho_C)-S(\rho)
\eeq
where 
$\rho=\rho_{ABC}$ and $S(\rho_j)=-tr[\rho_j \ln (\rho_j)]$ is the von Neumann entropy 
of the reduced state $\rho_j$. The classical correlation for the tripartite case is 
also generalized as
\beq
\mathcal{J}(\rho)=\max_{p\{i,j,k\}} \left [S(\rho_j)-
S(\rho_{j|i})+S(\rho_k)-S(\rho_{k|ij})\right]
\label{totclassicalcorre}
\eeq
where 
the maximum is taken over all the possible permutations of the indices $i$, $j$ 
and $k$. The conditional entropy is defined as $S(\rho_{j|i})=\min_{E^i_l}[S(j|\{E^i_l\})]$, 
where $S(j|\{E^i_l\})=\sum_l p_l S\left(r_i(E^i_l \rho_{ij} E^i_l / p_l ) \right)$. 
The probabilities $p_l$ are obtained as usual, $p_l = \,$Tr$_{ij}\left(\rho_{ij} E^i_l\right)$. 
The $E^i_l$  represents a set of POVMs for the parties $i$. 
An analogous definition for the conditional entropy is applied for the case 
in which two parties are measured $S(\rho_{k|ji})=\min_{E^i_l,E^j_m}[S(j|\{E^i_l,E^j_m\})]$. 
With this two generalizations it is possible to define the total discord for 
the system $ABC$, as
\beq
\mathcal{D}(\rho)=\mathcal{J}(\rho)-T(\rho).
\label{eq:TQD}
\eeq 
Now, is easy to see that the genuine quantum correlations can be obtained by 
a simple subtraction of the bipartite quantum corrections in the expression 
(\ref{eq:TQD}). So the GQD can be defined as
\beq
\mathcal{D}^{(3)}(\rho)=\mathcal{D}(\rho)- 
\mathcal{D}^{(2)}(\rho),
\eeq
where $\mathcal{D}^{(2)}(\rho)=\max[\mathcal{D}(\rho_{AB}),\mathcal{D}(\rho_{AC}),
\mathcal{D}(\rho_{BC})]$ being $\mathcal{D}(\rho_{ij})$ the 
usual definition of the QD for the bipartite system $ij$ 
\cite{ollivier02}:

\beq
D(\rho_{ij})=\min_{\{ E^i_l \}}\left[ S(\rho_i)- S(\rho_{ij})+S(j|\{E^i_l\})\right].
\eeq

\begin{figure}[tbp]
\centering 

  \includegraphics[bb=51 5 797 591,width=3.06in,height=2.45in,keepaspectratio]{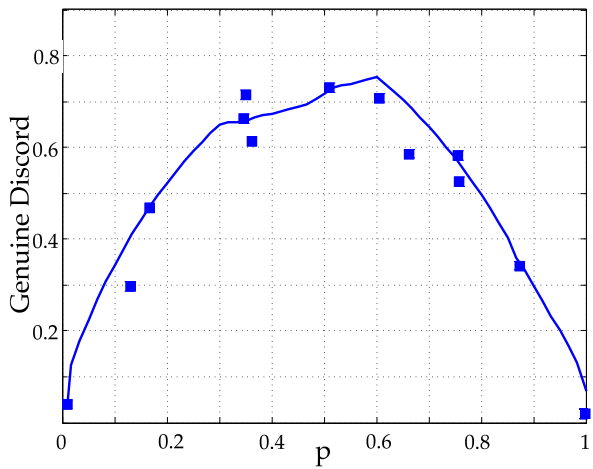}

\caption{(Color online) Evolution of the genuine Quantum Discord.
 Evolution for the AD channel (blue squares), and its theoretical evolution from the initial state
(blue line).}
\label{fig:fidelidadesmixed}
\end{figure}

Let us analyze the experimental data using these quantities. 
We begin by examining the dynamics for the measured states shown in Fig. \ref{fig:densidadampl}. 
In Fig. \ref{fig:fidelidadesmixed} we show the evolution of the GQD for the AD interaction. 
We can see that the experimental results (blue squares) are quite close to the evolution obtained
from the application of the theoretical map to the measured initial state (blue line). 
We can also see that the blue curve presents abrupt changes. 
It is related to the permutation of the terms in Eq. (\ref{totclassicalcorre}), 
which maximizes $\mathcal{J}$. This is conceptually different from the abrupt changes observed in the evolution of a bipartite system \cite{maziero09}, which occur when the optimal set of measurement operators $\{E^j_m\}$ changes during the evolution. In the tripartite case, the abrupt changes also depend on the partitions considered for the computation of the correlations.

For the PD channel (not shown), the experimental results cannot be described by the theoretical evolution of the initial state. The most probable reason is that the phase
damping channel is highly sensitive to phase fluctuations (even very small ones) in the interferometers, as we have already seen in the analysis of the entanglement. 

%%%%%%%%%%%%%%%%%%%%%%%
%%%%%%%%%%%%%%%%%%%%%%%

\begin{figure}[tbp]
\centering 

  \includegraphics[bb=43 9 795 591,width=3.06in,height=2.45in,keepaspectratio]{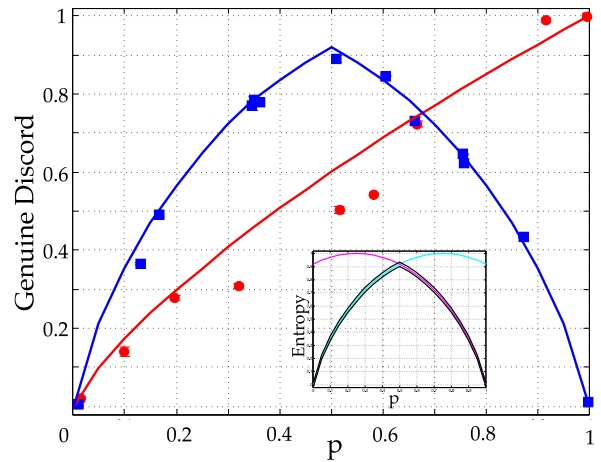}

\caption{(Color online) Evolution for the genuine Quantum Discord in the quasi-pure approximation.
Evolution for the AD channel (blue squares), and theory (blue solid line). 
Evolution for the PD channel (red circles) and theory (red line).
In the inset we plot the theoretical predictions for the reduced 
entropies: $S(\rho_E)$ in cyan and $S(\rho_B)$ in magenta.}
\label{fig:genuinediscord}
\end{figure}

Let us now analyze the same evolutions as before, using the dominant eigenvectors obtained in the spectral decomposition of Eq. (\ref{spec:decompose}), and shown in Fig. \ref{fig:autovalores}.  For pure states it was shown in Ref. \cite{giorgi11} that the  
GQD can be simply calculated as
\begin{equation}
\label{eq:pure_genui_discord}
\mathcal{D}^{(3)}=\min_i [S(\rho_i)],
\end{equation}
where $S$ is the Von Neumann entropy and $\rho_i$ is the reduced density matrix of the system $i$.  
In Fig. \ref{fig:genuinediscord} we can see the evolution
of the genuine QD for the two quantum channels.  
For the AD channel, there is a very good agreement between experimental data (blue squares) and 
theory (blue solid line). We can see a sudden change at $p=0.5$. 
This is related to the fact that the reduced entropy that takes the smallest value for $p<0.5$ is different from the one that takes the smallest value for $p>0.5$. 
In the inset, we show the evolution of the two reduced entropies of interest. 
For $p < 0.5$ $S(\rho_E)$ (environment reduced entropy) is the one that takes the smallest value among all the reduced entropies. This happens until $p=0.5$, point at which  $S(\rho_B)$ (system B reduced entropy) starts to take the smaller value. 
%%%%%%%%%%%%%%%%%%%%%%%%%%%%%%%%%%%%%%%%%%%
%%%%%%%%%%%%%%%%%%%%%%%%%%%%%%%%%%%%%%%%%%%%

The reduced entropy $S(\rho_i)$ is a measure of bipartite entanglement between system $i$ and system $jk$ for pure states \cite{vedral97}.  Thus, we can associate the abrupt change in GQD with the redistribution of entanglement in the tripartite system. In particular, we can observe that the minimum reduced entropy, corresponding to the minimum bipartite entanglement between system $i$ and system $jk$, is given by  $S (\rho_E)$ for $p < 0.5$ and by $S(\rho_B)$ for $p > 0.5$. In this sense, $p = 0.5$ appears as the value at which the transition between these two regions occurs.  

This effect can  be analyzed in terms of the monogamy relations. For pure states, we can replace 
$S (\rho_i)$ with $ C_{i (jk) }^{2}$ in Eq. ($\ref{eq:pure_genui_discord}$), and since $\tau_{ijk}=0$  
for the AD evolution, the  monogamy relation in Eq. (\ref{descomposicion}) can be rewritten as
\beq
\label{disc-entangl}
Ê \min_i[C _{i(jk)}^{2}] =\min_i [C_{ij}^{2} + C_{ik}^{2}],
\eeq
showing that the GQD can also be understood in terms of the bipartite entanglement between two subsystems.
Fig. \ref{fig:genuinediscord} shows that, for $p<0.5 $ the minimum tangle is given by $C_{E (AB)}^{2} $ which corresponds to the sum of the blue and red  curves in  Fig \ref{fig:conc_amp}. On the other hand, for $p> 0.5$ the minimum tangle is $ C_ {B (AE)} ^ {2} $ related with the sum of blue and black curves in Fig. \ref{fig:conc_amp}. In this context, we can observe that GQD discontinuity is related to the crossing between the two lines of Fig. \ref{fig:conc_amp}.

%%%%%%%%%%%%%%%%%%%%%%%%%%%%%%%%%%%
%%%%%%%%%%%%%%%%%%%%%%%%%%%%%%%

For the case of the PD channel, the experimental data (red circles) in this analysis have a reasonable agreement with the theory (red solid line). However, in the same way as in the analysis of the entanglement,
even in the quasi-pure approximation, the agreement between experiment and theory is not perfect. 
We notice that an advantage of the genuine discord, as compared to the entanglement parameter 
3-tangle, is that it is able to detect genuine quantum correlations for both tripartite families: 
the W-type and the GHZ-type \cite{giorgi11}.
This is clearly seen in Fig \ref{fig:genuinediscord}. 
%%%%%%%%%%%%%%%%%%%%%%%
%%%%%%%%%%%%%%%%%%%%%%%
  
\begin{figure}[tbp]
\centering 

  \includegraphics[bb=41 9 793 595,width=3.06in,height=2.45in,keepaspectratio]{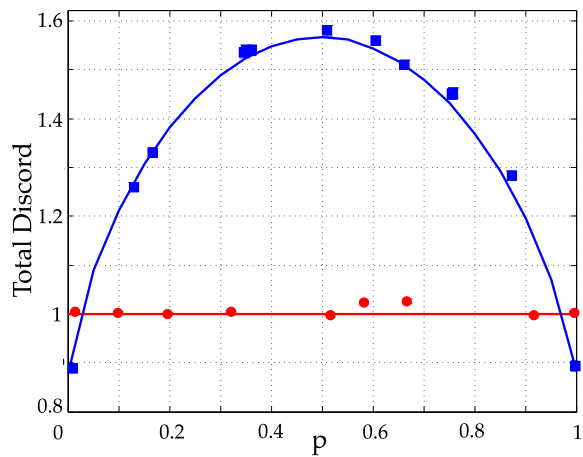}

\caption{(Color online) Evolution of the Total Quantum Discord.
Evolution for the AD channel (blue squares), and theory (blue line). 
Evolution for the PD channel (red circles) and theory (red line).}
\label{fig:disctotal}
\end{figure}

Another interesting aspect of this approach is that it permits to analyze the non-classicality of the 
tripartite states for each dynamics. 
This can be done with the total Quantum Discord (TQD) defined in  \cite{giorgi11}. 
In Fig \ref{fig:disctotal} we show the TQD computed from the pure states obtained from the
the spectral decomposition of the measured states, as described in Sec. \ref{sec:decomp}. 
We can see that during the evolution of the AD channel (blue squares), the
TQD takes greater values than the ones corresponding to the evolution for the PD channel (red circles). 
This is related to the fact that the W-family possesses not only genuine tripartite entanglement, but also  bipartite entanglement. We can observe that the TQD for the case of PD is constant during all the evolution, like the invariant shown in Fig. \ref{fig:tanglesdeph}. This means that 
there are also changes in the kind of correlation along the evolution. 
In the beginning it is completely bipartite, and in the end it is tripartite. 

The fact that some states have higher TQD indicates that there might exist some task, like the one reported in \cite{dakic12}, for which these states perform better. 

%%%%%%%%%%%%%%%%%%%%%%%
%%%%%%%%%%%%%%%%%%%%%%%

%%%%%%%%%%%%%%%%%%%%%%%%%%%%%%%%%%%%%%%%%%%%%%%%%%%%%%%%%%%%%
\section{Conclusions}

\label{sec:conclusions}

We presented in this paper a detailed theoretical and experimental analysis of the flow of quantum correlations, including entanglement and quantum discord, for an initially entangled state of two qubits coupled to a local environment. A recently proposed experimental photonic setup is the perfect scenario for this study, since it allows full tomography of the entangled system and its environment. The main purpose of our experimental investigation was to get new insights into the process of decay of entanglement in open-system dynamics, and into the emergence of genuine multipartite quantum correlations between a system and its environment. 

The mechanism of distribution and mutation of bipartite entanglement was elucidated with the help of monogamy relations for open systems. The emergence of GHZ entanglement was analyzed with the help of a useful expression for the 3-tangle, which is shown to be expressible  as a product of the initial entanglement and a function of solely the Kraus operators that define the open-system dynamics. This result is valid for a wide class of important quantum channels.

On the other hand, the emergence of W-type entanglement was signaled by the appearance of genuine multipartite quantum discord, which was investigated experimentally here for the first time. The dynamics of this quantity may exhibit a new non-analytical behavior, which was shown to occur at the same instant of time when the genuine W-type entanglement between the two-qubit system and the environment becomes maximal. We emphasized that this non analytical behavior is different from other similar phenomena already reported in the literature for bipartite systems, and has its origin in the distribution of correlations in the different partitions of the tripartite system. 

Our experiment aims to emulate the isolated dynamics which comes from the inclusion of the environment. While our results imply global states with very high purity, ideal  pure states are of course never reconstructed. Nevertheless, by considering the main component in the spectral decomposition of the density matrix, we were able to conciliate our experimental results  with the theory developed for pure states. This technique, quite useful in the present context, may be easily generalized to other systems involving the dynamics of quasi-pure states.

%%%%%%%%%%%%%%%%

\begin{acknowledgements}
Financial support was provided by Brazilian agencies CNPq, CAPES,
FAPERJ, and the National Institute of Science and Technology for Quantum Information. 
OJF and SPW acknowledge funding from the FET-Open Program, within the 7th Framework Programme of
the European Commission under Grant No. 255914 (PHORBITECH).
A.V.H. was funded by the Consejo Nacional de Ciencia y Tecnolog\'{\i}a, M\'{e}xico.
\end{acknowledgements}

%%%%%%%%%%%%%%%%%%%%%BIBLIOGRAPHY%%%%%%%%%%%%%%%%%%%%%%%%%%%%%%%

%\bibliography{Master_Bibtex}
%\bibliographystyle{apsrev}

%\bibliography{/Users/spwalborn/Dropbox//Master_Bibtex}
%\bibliography{multipartite}
%\bibliography{multipartite_gabo}
%%%%%%%%%%%%%%%%%%%%%%%%%%%%%%%%%%%%%%%%%%%%%%%%%%%%%%%%%%%%%%%%%%

\end{document}